\documentstyle{mn}
\input{epsf}
\begin{document}

\def\ao{\alpha_{\rm 2-18}}
\def\ee{$e^\pm$}
\def\g{$\gamma$}
\def\nh{N_{\rm H}}
\def\af{A_{\rm Fe}}
\def\taut{\tau_{\rm T}}
\def\exosat{{\it EXOSAT}}
\def\ginga{{\it Ginga}}
\def\heao{{\it HEAO-1}}
\def\ec{E_{\rm c}}
\def\efe{E_{\rm Fe}}
\def\sfe{\sigma_{\rm Fe}}
\def\ife{I_{\rm Fe}}
\def\loc{\ell_{\rm l}}

\newbox\grsign \setbox\grsign=\hbox{$>$} \newdimen\grdimen \grdimen=\ht\grsign
\newbox\simlessbox \newbox\simgreatbox \newbox\simpropbox
\setbox\simgreatbox=\hbox{\raise.5ex\hbox{$>$}\llap
     {\lower.5ex\hbox{$\sim$}}}\ht1=\grdimen\dp1=0pt
\setbox\simlessbox=\hbox{\raise.5ex\hbox{$<$}\llap
     {\lower.5ex\hbox{$\sim$}}}\ht2=\grdimen\dp2=0pt
\setbox\simpropbox=\hbox{\raise.5ex\hbox{$\propto$}\llap
     {\lower.5ex\hbox{$\sim$}}}\ht2=\grdimen\dp2=0pt
\def\simgreat{\mathrel{\copy\simgreatbox}}
\def\simless{\mathrel{\copy\simlessbox}}

\topmargin = -1cm

\title[X-ray/gamma-ray spectrum of Seyfert 1s]
{The average X-ray/gamma-ray spectrum of radio-quiet Seyfert 1s}
\author[D. Gondek et al.]
{\parbox[]{6.5in} {Dorota Gondek$^1$, Andrzej A. Zdziarski$^{1,2}$, W. Neil 
Johnson$^3$, Ian M. George$^{4,5}$, Kellie McNaron-Brown$^6$, 
Pawe\l\ Magdziarz$^7$, David Smith$^8$ and Duane E. Gruber$^9$} \\
$^1$N. Copernicus Astronomical Center, Bartycka 18, 00-716 Warsaw, Poland; 
Internet: (dorota, aaz)@camk.edu.pl \\
$^2$Institute for Theoretical Physics, University of California, Santa Barbara,
CA 93106, USA \\
$^3$E. O. Hulburt Center for Space Research,
Naval Research Laboratory, Washington, DC 20375, USA \\
$^4$Laboratory for High Energy Astrophysics, Code 660.2, 
NASA/Goddard Space Flight Center, Greenbelt, MD 20771, USA \\
$^5$Universities Space Research Association, USA \\ 
$^6$George Mason University, Fairfax, VA 22030, USA \\
$^7$Astronomical Observatory, Jagiellonian University, Orla 171, 30-244 Cracow, 
Poland\\
$^8$Department of Physics, University of Leicester, University Road, Leicester 
LE1 7RH, UK\\
$^9$Center for Astrophysics and Space Science, University of California at San
Diego, La Jolla, CA 92093, USA} 

\maketitle

\begin{abstract}

We have obtained the average 1--500 keV spectrum of radio-quiet Seyfert 1s 
using data from \exosat, \ginga, \heao, and {\it GRO\/} OSSE. The spectral fit 
to the combined average \exosat\/ and OSSE data is fully consistent with that 
for \ginga\/ and OSSE, confirming results from an earlier \ginga/OSSE sample. 
The average spectrum is well-fitted by a power-law X-ray continuum with an 
energy spectral index of $\alpha \simeq 0.9$ moderately absorbed by an ionized 
medium and with a Compton reflection component. A high-energy cutoff (or a 
break) in the the power-law component at a few hundred keV or more is required 
by the data. We also show that the corresponding average spectrum from \heao\/ 
A1 and A4 is fully compatible with that obtained from \exosat, \ginga\/ and 
OSSE. These results confirm that the apparent discrepancy between the results 
of \ginga\/ (with $\alpha \simeq 0.9$) and the previous results of \exosat\/ 
and \heao\/ (with $\alpha \simeq 0.7$) is indeed due to ionized absorption and 
Compton reflection first taken into account for \ginga\/ but not for the 
previous missions. Also, our results confirm that the Seyfert-1 spectra are on 
average cut off in \g-rays at energies of at least a few hundred keV, {\it 
not\/} at $\sim 40$ keV (as suggested earlier by OSSE data alone). The 
average spectrum is compatible with emission from either an optically-thin  
relativistic thermal plasma in a disk corona, or with a nonthermal plasma with 
a power-law injection of relativistic electrons. 

\end{abstract}


\section{INTRODUCTION}
\label{s:intro}

The average X-ray/\g-ray (hereafter X\g) spectra of Seyfert 1s and 2s observed 
by both \ginga\/ and {\it GRO\/} OSSE have recently been obtained by Zdziarski 
et al.\ (1995, hereafter Z95). The main result of that study for Seyfert 1s is 
that their average spectral high-energy cutoff is around several hundred keV. 
This is thus similar to the cutoff of IC 4329A, a bright Seyfert 1 (Madejski 
et al.\ 1995; Zdziarski et al.\ 1994, hereafter Z94). Z95 also found that the 
average spectrum of radio-quiet (hereafter RQ) Seyfert 1s has the energy 
spectral index of $\alpha\simeq 0.9$ as well as a Compton-reflection 
component, as previously obtained for \ginga\/ spectra alone (Pounds et al.\ 
1990; Nandra \& Pounds 1994, hereafter NP94). However, since the 23 \ginga\/ 
and 8 OSSE observations of RQ Seyfert 1s used in Z95 are not simultaneous and 
that sample consists of only 4 objects, there is a clear need to confirm those 
results using enlarged and independent samples. 

Here, we test and confirm the results of Z95 using spectra of RQ Seyfert 1s 
from \exosat, \heao, \ginga, and OSSE. [Discussion of the average X\g\ 
properties of radio-loud Seyferts is given in Wo\'zniak et al.\ (1996).] We 
obtain the average spectrum of RQ Seyfert 1s detected by both \exosat\/ and 
OSSE, a sample which consists of 7 objects. The objects were observed 41 times 
by \exosat\/ and 18 times by OSSE. Averaging this large numbers of 
observations is expected to compensate for the lack of simultaneity of the 
observations. We also obtain the corresponding average spectrum from \heao\/ 
A1 and A4. Furthermore, we analyze the \ginga/OSSE sample enlarged by new OSSE 
observations of NGC 5548 and with addition of NGC 7469. 
 
After presenting the spectra, we study physical processes that can be 
responsible for the observed X\g\ emission. We consider both thermal and 
nonthermal plasmas. 

\section{THE DATA}
\label{s:data}

We use \exosat\/ spectra from the HEASARC archive with the quality flag 3 or 
higher (which indicates observations with relatively reliable background 
subtraction) of RQ Seyfert 1s detected by OSSE. We exclude the Seyfert 1s 
brightest in X-rays, NGC 4151 and IC 4329A, in order to avoid their dominance 
of the co-added spectrum. The usable \exosat\/ energy range is from 1.2 keV to 
8 keV (channels 6--31). The spectra above 8 keV suffer from relatively 
inaccurate global background subtraction [as opposed to local background 
subtraction used by, e.g., Turner \& Pounds (1989)]. The individual spectra 
include a 1 per cent systematic error. The spectra from \exosat\/ (as well as 
\ginga\/ and \heao) are co-added with the weights corresponding to the length 
of time of each observation. Both the counts and the response matrices for 
each \exosat\/ observation are added using a procedure (specially written for 
\exosat\/ data) in the {\sc ftools} data processing package. 

The OSSE spectra take into account an estimated systematic error correction to 
the spectra [see, e.g., Zdziarski, Johnson \& Magdziarz (1996) for details]. 
We use the OSSE response matrix as revised in 1995, which results in the 
50--60 keV fluxes about 20 per cent higher than in the earlier response (used, 
e.g., in Z95).  

The \exosat/OSSE sample consists of 7 RQ Seyfert 1s [41, 18] (the numbers in 
brackets give the number of \exosat\/ and OSSE observations, respectively): MCG 
--6-30-15 [4, 2], Mrk 509 [3, 2], NGC 3783 [4, 1], NGC 5548 [11, 6], MCG 
8-11-11 [10, 2], ESO 141-55 [2, 2], and NGC 7469 [7, 3]. 

We also use the average spectrum in the 2--180 keV range from observations by 
\heao\/ A1 and A4 (Wood et al.\ 1984; Rothschild et al.\ 1983). [We do not 
use \heao\/ A2 data because the normalization of archival spectra is not 
known (Weaver, Arnaud \& Mushotzky 1995).] The A4 spectra of individual 
objects have been recreated using the current version of the instrument 
software. In the 2--10 keV energy range, we use the \heao\/ A1 data as 
published by Wood et al.\ (1984). Since that detector provides no spectral 
information and only instrumental counts are given in Wood et al.\ (1984), we 
have obtained the 2--10 keV fluxes using the counts-to-flux conversion as 
given for the Crab. In order to get an estimate of the 2--10 keV average 
spectrum, we use the average overall slope in that range of $\alpha=0.7$ (see 
Section \ref{s:rq} below). 

Our \ginga\/ data are the same as those for Seyfert 1s in Z95 except for the 
addition of NGC 7469 (NP94). The \ginga/OSSE sample thus consists of 5 RQ 
Seyfert 1s [25, 14] (the numbers in brackets give the number of \ginga\/ and 
OSSE observations, respectively): MCG --6-30-15 [4, 2], Mrk 509 [4, 2], NGC 
3783 [1, 1], NGC 5548 [14, 6], and NGC 7469 [2, 3]. We have obtained the 
average spectra for both the top-layer and the mid-layer \ginga\ data (Turner 
et al.\ 1989). However, we have found that the mid-layer spectrum above 10 keV 
is systematically softer than the corresponding top-layer spectrum. Since the 
mid-layer calibration is much more uncertain than that of the top layer 
(Turner et al.\ 1989), we use in this paper only the top-layer data. We use 
the energy range of 1.7--18 keV (channels 4--31), for which the instrumental 
background subtraction is accurate. As in Z95, a 0.5 per cent  systematic 
error correction has been applied to the co-added \ginga\/ spectrum. 

\section{MODEL}
\label{s:model}

In our fits, we use the {\sc xspec} spectral fitting package version 9.0 
(Arnaud 1996). As in Z95, we model the underlying continuum as a power law 
with an exponential cutoff at an energy, $\ec$. Z95 found that ionized 
low-energy absorption is necessary to explain the average \ginga\/ spectrum of 
Seyfert 1s. Thus, we model absorption as due to an ionized medium with the 
abundances from Anders \& Ebihara (1982) and the ion opacities of Reilman \& 
Manson (1979) except for the Fe K-edge energies, for which results of Kaastra 
\& Mewe (1993) were used. The absorber temperature is assumed to equal 
$T=10^5$ K (Krolik \& Kallman 1984), its column density is $N_{\rm H}$, 
and an ionization parameter is defined by $\xi=L/(nr^2)$. Here $L$ is the 5 
eV--20 keV luminosity in a power law spectrum with the average energy index of 
0.7, and $n$ is the density of the absorber located at distance $r$ from the 
illuminating source. Model parameters are given at the average redshift for 
each sample, $\langle z\rangle$. In addition to the ionized absorber, we 
include a fixed neutral absorber at $z=0$ with $N_{\rm H,G}$ equal to the 
average Galactic value for each sample. 

The underlying continuum irradiates cold matter, e.g., an accretion disk, and 
gives rise to a Compton-reflection spectral component (Lightman \& White 
1988). Differently from Z95, who used the reflection spectrum averaged over 
all the angles of the reflected photons (Lightman \& White 1988), we assume 
a viewing angle of $30^\circ$ [corresponding to an orientation close to 
face-on expected in type-1 AGNs, Antonucci (1993)]. We use the angle-dependent 
reflection Green's functions of Magdziarz \& Zdziarski (1995). The luminosity 
intercepted by the reflecting medium equals $R$ (the relative contribution of 
reflection) times the luminosity emitted outward, i.e., $2\pi R$ gives the 
solid angle subtended by the absorber as seen from the source of the incident 
radiation. The continuum reflection is accompanied by emission of a 
fluorescent Fe K$\alpha$ line, which we model here as a Gaussian at an energy 
$\efe$ and the width $\sfe$.

\section{RESULTS}
\label{s:rq}

We first study the \ginga/OSSE average spectrum, using the sample enlarged 
with respect to that in Z95 (see Section \ref{s:data}). The data and the 
best-fit model are shown in Fig.\ 1{\it a}. The fit parameters are given in 
Table 1. We obtain $\alpha\simeq 0.90\pm 0.05$ and the $e$-folding energy 
between $\sim 0.4$ MeV and 2.7 MeV, which is in agreement with Z95. The 
relative contribution of reflection, $R\simeq 0.76\pm 0.15$ is about 2/3 of 
that found by Z95, which is explained by the angle-dependent reflection 
spectrum at $30^\circ$ both having a higher normalization with respect to the 
underlying continuum and being harder in the $\sim 10$--30 keV range than the 
angle-averaged reflection spectrum (Magdziarz \& Zdziarski 1995). [Note that 
this effect will also reduce large values of angle-averaged $R$ obtained by 
Weaver et al.\ (1995) for Seyfert 1s observed by \heao\/ A2.]  Our values of 
$\alpha$ and $R$ are in agreement with the average values for \ginga\/ of 
NP94, $\alpha\simeq 0.95$ with a dispersion of 0.15 and $R\sim 0.5$--0.7. The 
equivalent width of the K$\alpha$ line is $120_{-40}^{+40}$ eV, which agrees 
with the value expected from fluorescence in the reflecting medium at our $R$ 
(George \& Fabian 1991). We have also confirmed that our results are only 
weakly dependent on the (relatively uncertain) elemental abundances. When the 
abundances of Anders \& Ebihara (1982) are replaced by those of Anders \& 
Grevesse (1989) (with higher abundances of metals, in particular with about 40 
per cent more Fe), our results change only slightly. E.g., marginally more 
Compton reflection is obtained, $R= 0.80^{+0.16}_{-0.14}$ ($\chi^2 = 71.8/72$ 
dof). 

\begin{figure}
\begin{center}
\leavevmode
\epsfxsize=8.4cm \epsfbox{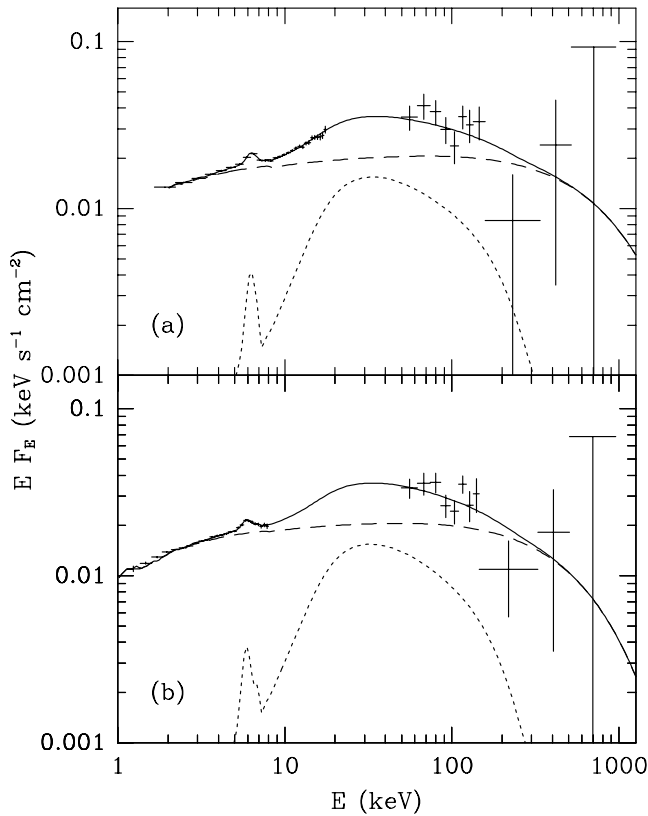}
\end{center}
\label{fig:rq}
\caption{
The average spectrum ({\it crosses}) of RQ Seyfert 1 galaxies from \ginga/OSSE 
{\it (a)\/} and \exosat/OSSE {\it (b)}. The upper limits here and in figures 
below are 2-$\sigma$. The dashed curves represent power-law spectra with 
exponential cutoffs and bound-free absorbed at low energies, and the dotted 
curves represent Compton reflection including the Fe K$\alpha$ line. The solid 
curves give the sum. } 
 \end{figure}

\begin{table*}
\label{t:par}
\centering
\caption{Parameters of the spectral fits to the average RQ Seyfert 1 spectrum. 
The \ginga/OSSE and \exosat/OSSE samples are denoted by `GO' and `EO', 
respectively. The symbols are explained in the text. Parameters with no
error ranges are fixed. $\nh$ are in units of $10^{21}$ cm$^{-2}$, $\ec$, 
$E_{\rm Fe}$, and $\sigma_{\rm Fe}$ are in keV. Errors are for 90 per cent 
confidence limit based on a $\Delta\chi^2=2.7$ criterion (Lampton, Margon \& 
Bowyer 1976). 
 } 
\begin{tabular}{lcccccccccc}
\hline
Data & $\langle z\rangle$ & $\alpha$ & $\ec$ & $R$ &  $N_{\rm 
H,G}$ & $N_{\rm H}$ & $\xi$ & $E_{\rm Fe}$ &
$\sigma_{\rm Fe}$ & $\chi^2$/dof\\
GO &
0.017 & $0.90^{+0.05}_{-0.05}$ & $730^{+1950}_{-340}$ & $0.76^{+0.15}_{-0.15}$ 
&0.5 & $32^{+19}_{-19}$  & $530^{+310}_{-300}$ & $6.32^{+0.15}_{-0.16}$ & 
$0.33^{+0.28}_{-0.33}$ & 74.1/72\\
EO &
0.020 & $0.90^{+0.09}_{-0.07}$ & $510^{+4300}_{-250}$ & 0.76 & 0.7 &
$16^{+24}_{-14}$ & $140^{+340}_{-140}$ & $6.0^{+0.2}_{-0.2}$ &
$0.3^{+0.5}_{-0.3}$ & 73.3/67\\
\hline
\end{tabular}
\end{table*}

Then we consider the \exosat/OSSE sample, see Fig.\ 1{\it b}. Since 
the usable \exosat\/ spectrum extends only to 8 keV, Compton reflection is not 
constrained. Therefore, we fix $R$ at the value obtained from the \ginga/OSSE 
average. The equivalent width of the K$\alpha$ line obtained, $100^{+100}_{-
50}$ eV, is consistent with this assumption. (The best-fit line energy is 5 
per cent less than that for \ginga, which is due to a gain inaccuracy of 
\exosat.) Furthermore, we find that there is an apparent soft X-ray excess 
present in the average spectrum below 2 keV, a feature common in Seyfert 1 
spectra (Wilkes \& Elvis 1987; Turner \& Pounds 1988). Rather than add a 
separate spectral component (which would be poorly constrained by our data), 
we fit only the spectrum above 2 keV. We show, however, the spectrum below 2 
keV in Fig.\ 1{\it b}.  

We find that the fit parameters of the \exosat/OSSE average spectrum are 
virtually identical to those for the \ginga/OSSE average. The two spectra are 
compared in Fig.\ 2. We thus confirm that the average value of the $e$-folding 
energy in the spectra of Seyfert 1s is several hundred keV rather than 
$\sim 40$ keV (Johnson et al.\ 1994). The latter value was obtained using 
narrow-band OSSE data only, neglecting Compton reflection, using a spectral 
model implying X-ray power laws much harder than observed, as well as with the 
old OSSE response. Those factors fully explain the discrepancy. 

\begin{figure*}
\label{fig:comp}
\begin{center}
\leavevmode
\epsfxsize=12cm \epsfbox{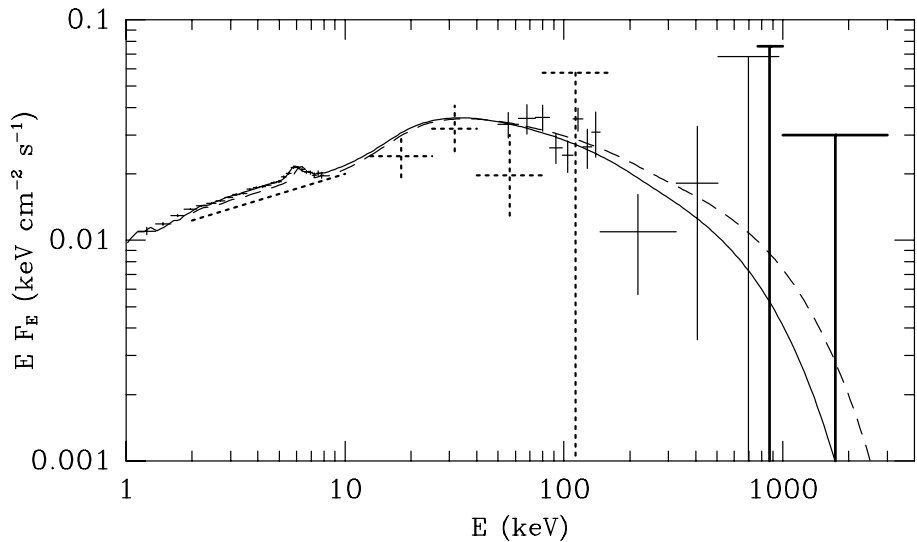}
\end{center}
\caption{
Comparison of the estimates of the average spectrum of RQ Seyfert 1s obtained 
with various instruments. The solid-line data are from \exosat\/ and OSSE. The 
solid and dashed curves are the best-fit models for the \exosat/OSSE and 
\ginga/OSSE samples, respectively. Note that the two curves have the actual 
normalizations from the fits. The agreement between the two spectra confirms 
that the X-ray variability of the AGNs is indeed fully compensated by the 
sizes of the samples. The dotted symbols represent the average \heao\/ A1 and 
A4 spectrum, which is in agreement with the \ginga/\exosat/OSSE spectrum. The 
thick-line upper limits are for combined emission of all Seyfert galaxies 
observed by COMPTEL (Maisack et al.\ 1995).} 
 \end{figure*}

We point out that the OSSE spectra of individual Seyferts appear more uniform 
(Johnson et al.\ 1994) than their X-ray spectra, for which the 1-$\sigma$ 
dispersion of $\alpha$ is $0.15\pm 0.04$ (NP94).  If indeed AGNs with 
different $\alpha$ have similar \g-ray spectra, there will be a positive 
correlation of $\ec$ with $\alpha$, rather than a constancy of $\ec$ among 
objects with different $\alpha$ (because harder X-ray spectra will need to be 
cut off stronger than softer ones in order to yield similar spectra above 50 
keV). The range of $\ec$ among individual AGNs will be then larger than that 
given in Table 1, which range corresponds to our {\it average\/} X-ray 
spectrum, with $\alpha=0.90\pm 0.05$. This appears to be confirmed for NGC 
4151, a bright Seyfert with a hard X-ray spectrum. Zdziarski et al.\ (1996) 
have found no statistical difference between the shape of the spectra of NGC 
4151 from four OSSE observations during 1991--93 and the average OSSE spectrum 
of the Seyfert 1s observed by \exosat. However, the $e$-folding energy for NGC 
4151 is $\ec\simeq 150$ keV (below the range of $\ec$ in Table 1) as a 
consequence of $\alpha\simeq 0.7$ [for the cut-off power-law model with 
reflection fitted to 1991 June \ginga/OSSE observation (Zdziarski et al.\ 
1996)]. 

The agreement between the average \exosat\/ and \ginga\/ spectra confirms that 
the average value of $\alpha\simeq 0.7$ for Seyferts found based on \exosat\/ 
observations (Turner \& Pounds 1989) is indeed an artifact of assuming neutral 
absorption and not including Compton reflection. When both effects are 
included, the average $\alpha\simeq 0.9$ for both \exosat\/ and \ginga. 

A similar average spectral index of $\alpha\simeq 0.6$--0.7 was also obtained 
by \heao\/ (Rothschild et al.\ 1983; Mushotzky 1984). Here, we obtained the 
average \heao\/ A1/A4 spectrum for the objects in the \exosat/OSSE sample (see 
Section \ref{s:data}), shown as dotted symbols in Fig.\ 2. We see that apart 
from a small difference in the normalization the two spectra are fully 
consistent with each other. Thus, the spectra of Seyfert 1s from \exosat\/ and 
\heao\/ {\it are\/} compatible with an intrinsic power law with $\alpha\simeq 
0.9$. Our average \heao\/ spectrum is almost the same as that obtained for a 
larger sample of Seyfert 1s by Maisack, Wood \& Gruber (1994). 

Fig.\ 2 also shows the upper limits on emission of {\it all\/} Seyferts 
observed by COMPTEL (Maisack et al.\ 1995). The corresponding upper limits for 
the \exosat/OSSE sample of RQ Seyfert 1s are about 3--4 times higher. We see 
that even the limits for all Sefyerts do not provide additional constraints on 
our average spectrum. The upper limits on \g-ray emission of Seyferts were 
also obtained by EGRET for energies above 100 MeV (Lin et al.\ 1993). The 
weighted  2-$\sigma$ upper limit for the \exosat/OSSE sample corresponds to 
$EF_E \simless 0.004$ keV cm$^{-2}$ s$^{-1}$ at 100 MeV. This is much above 
our extrapolated models. However, that upper limit does rule out an 
extrapolation of the $\alpha=0.9$ power law  {\it without\/} a break up to 100 
MeV, at which energy the extrapolated power law spectrum would be a factor of 
$\sim 5$ above the upper limit. Thus, the EGRET results provide a confirmation 
of the existence of a high-energy spectral break in Seyferts, independent of 
the OSSE results. 

\section{PHYSICAL PROCESSES IN SEYFERTS}

\subsection{Thermal models}

Exponentially cut-off power-laws (used here to fit the average Seyfert-1 
spectrum) can be used to model spectra due to Comptonization in thermal, 
optically-thin, mildly-relativistic plasmas (e.g., Z94). The parameters,
$\alpha$ and $\ec$, can be related to the Thomson optical depth, $\tau$, 
and temperature, $T$. For $\ec=400$ keV, which fits both our average spectra 
and IC 4329A, Z94 find $\ec\simeq 1.6kT$ (implying $kT\simeq 260$ keV). Z94 
also provide an expression for $\alpha(\tau, kT)$ in a slab geometry, which 
yields $\tau\simeq 0.1$ for $\alpha=0.95$, $\ec=400$. Thus, optically-thin, 
mildly relativistic plasmas can explain the observed spectra of Seyfert 1s. 
The large values of $\ec$ obtained here rule out models with Comptonization in 
optically thick plasmas (Sunyaev \& Titarchuk 1980), used in the past to fit 
Seyfert-1 spectra (e.g., Miyoshi et al.\ 1988). 

Haardt \& Maraschi (1993) have proposed that the hot plasma in Seyferts forms 
a corona above the surface of an accretion disk, and that most of the energy 
dissipation occurs in the corona. Then the plasma temperature can be 
determined from the disk-corona energy balance, which makes the model more 
self-consistent. The disk-corona model also accounts for the 
Compton-reflection spectral components in the spectra of RQ Seyfert 1s. Z94 
found that the best-fit parameters of the X\g\ spectrum of IC 4329A satisfy 
that energy balance, i.e., the hard corona emission reprocessed by the disk 
self-consistently provides the seed of soft photons (in the UV range) for 
Compton upscattering into the hard spectrum.  

Those results have recently been confirmed by a more sophisticated treatment of 
the radiative transfer in Stern et al.\ (1995; hereafter S95). We apply 
their results on homogeneous slab coronae to our average spectrum of RQ Seyfert 
1s. The spectrum has $\alpha\simeq 0.9$ and the overall 2--18 keV spectral 
index, $\ao\simeq 0.7$ (including the reflection component). For that $\ao$ 
and assuming a pure \ee\ corona, S95 obtain $\tau \simeq 0.05$ and $kT\simeq 
330$ keV (roughly corresponding to $\ec\simeq 500$ keV) in agreement with 
$\ec$ obtained from our fits, see Table 1. This agreement supports the 
dissipative corona model of Haardt \& Maraschi (1993).

On the other hand, Haardt, Maraschi \& Ghisellini (1994) point out that 
since the UV fluxes in many Seyfert 1s are much larger than the X-ray fluxes 
(Walter \& Fink 1993) the model of Haardt \& Maraschi (1993), with most of the 
dissipation occuring in the corona, is ruled out. In that model, the UV 
emission is due to reprocessing of the X\g\ emission directed towards the cold 
disk, and $EF_E$ in the UV and in X-rays are expected to be of the same order 
of magnitude. To the contrary, $EF_E(1375\,{\rm \AA})/EF_E(2\,{\rm keV})$ 
[from the best-fit values in Walter \& Fink (1993)] for our RQ sample equals 
0.8, 36, 0.2, 4.2, 19, 17, and 12 for MCG 8-11-11, NGC 3783, MCG --6-30-15, 
NGC 5548, ESO 141-G55, NGC 7469, and Mrk 509, respectively. We see that the 
ratio is $> 10$ for 4 objects. These large ratios can be explained if the 
corona is patchy rather than homogeneous and the corona dissipation dominates 
the disk dissipation only in the vicinity of an active region (a `patch'), but 
not globally (Haardt et al.\ 1994). 

We also point out that a pure \ee\ pair corona would not form a thin slab 
above the disk surface (which was assumed by Haardt \& Maraschi 1993). 
Hydrostatic equilibrium implies $H_{\rm c}/R_{\rm d} =(2r\Theta)^{1/2}$ for a 
pure pair, gas pressure-dominated, corona, where $H_{\rm c}$ is the corona 
scale-height, $R_{\rm d}$ is the disk radius, $r\equiv R_{\rm d}c^2/(2GM)$, 
and $\Theta \equiv kT/(m_e c^2)$. Thus, $H_{\rm c}/R_{\rm d} \simgreat 1$ for 
$\Theta r\simgreat 1$ and the assumption of a slab geometry breaks down. 

S95 also provide one more argument against homogeneous slab coronae based on 
pair equilibrium. They calculate the maximum local compactness, $\loc \equiv 
L_{\rm l} \sigma_{\rm T}/(H_{\rm c} m_{\rm e} c^3)$, at which the homogeneous 
corona can be in pair equilibrium. Here $L_{\rm l}$ is the luminosity from a 
characteristic local volume, $H_{\rm c}^3$, in the corona. For $\ao \simeq 
0.7$ of the \ginga/OSSE sample, S95 obtain $\loc\simless 0.4$. The maximum 
value is achieved in a pure pair corona, and a presence of ionization 
electrons lowers the equilibrium value of $\loc$. S95 point out that this 
is in conflict with the AGN variability data (Done \& Fabian 1989). On the 
other hand, S95 show that higher $\loc$ are possible for patchy coronae. For 
$\ao\simeq 0.7$ of the \ginga/OSSE sample, $\loc\simless 15$ for an active 
region in the form of a hemisphere. That compactness appears compatible with 
the compactnesses estimated from X-ray variability (Done \& Fabian 1989). The 
plasma parameters of RQ Seyfert 1s implied by thermal Comptonization are 
$\tau\simeq 0.2$ and $kT\simeq 0.7$ MeV in the case of pure pairs (see Fig.\ 1 
in S95), which $kT$ is compatible with our fitted values of $\ec$. 

However, we point out that the results of S95 need to be modified to account 
for the difference between the local compactness (used to compute pair 
equilibria) and the global compactness, $\ell \equiv L \sigma_{\rm T}/ (R_{\rm 
d} m_{\rm e} c^3)$ (where $R_{\rm d}$ is the radius of the region where the 
X\g\ luminosity $L$ is produced). The latter (rather than the former) 
compactness is constrained from the variability data by $R_{\rm d}\simless 
c\Delta t$ (Done \& Fabian 1989). Using the formalism of Svensson \& Zdziarski 
(1994) for the scale-height of the corona, assuming $\Theta\simeq 0.5$, and 
using the constraint on the size of the X-ray producing region of $r\simless 
20$ (Tanaka et al.\ 1995; Fabian et al.\ 1994; Mushotzky et al.\ 1995), we 
obtain $\ell\simgreat 40 (1+n_+/n_{\rm p})^{-1/2}\loc$. Thus, $\ell\gg \loc$ 
is possible in coronae. This largely resolves the conflict between the small 
local compactness required by pair equilibrium in a homogeneous corona and the 
variability data. Thus, the main argument against homogeneous coronae remains 
the large UV fluxes (Haardt et al.\ 1994), rather than pair production (S95). 

More evidence against homogeneous coronae in Seyfert 1s is, however, provided 
by X-ray variability. Czerny \& Lehto (1996) find that some variability 
light-curves from \exosat\/ are truly stochastic rather than due to a 
deterministic chaos, which implies X-ray emission by multiple active centers 
rather than by a single extended source. 

We note that studies of thermal pair plasmas in pair equilibrium predict no 
distinct pair annihilation even from a pair-dominated plasmas 
(Macio{\l}ek-Nied\'zwiecki, Zdziarski \& Coppi 1995). A future detection of 
such a feature would indicate the plasma is either nonthermal 
with a hard electron injection (see Section \ref{ss:nth}) or there is a strong 
pair wind and pair annihilation takes place in a region spatially different 
from the hot corona (e.g., Macio{\l}ek-Nied\'zwiecki et al.\ 1995). 

\subsection{Nonthermal models} 
\label{ss:nth}

In nonthermal models, electrons are accelerated to or injected with a 
nonthermal distribution extending to relativistic energies. The electrons 
Compton upscatter UV seed photons to the X\g\ energy range, and the \g-rays 
may produce relativistic pairs, supplementing the primary injection of 
nonthermal electrons. We use here a numerical model of Lightman \& Zdziarski 
(1987) with modifications given in Zdziarski, Coppi \& Lamb (1990).  We also 
take into account Compton reflection and absorption, which best-fit parameters 
for models below are almost the same as for the cut-off power-law model, 
Section \ref{s:rq}. We fit the \ginga/OSSE data only since they provide more 
stringent constraints on the continuum parameters than the \exosat/OSSE data 
(see Section \ref{s:rq}). 

The simplest nonthermal model consists  of power law electrons with the 
steady-state index of $p=2\alpha+1\simeq 2.8$ that singly-scatter some soft 
seed photons in the Thomson regime, yielding a power-law photon distribution 
with $\alpha\simeq 0.9$ and a high-energy cutoff at $E\gg 511$ keV (Blumenthal 
\& Gould 1970). If the relativistic electrons lose energy in the Thomson 
regime and thermalize within the source, the index, $p$, corresponds to 
injection of relativistic electrons with an index less by one (Blumenthal \& 
Gould 1970), $\Gamma\simeq 2\alpha\simeq 1.8$. Since the model gives no 
high-energy cutoff in the hundred-keV range, it fits our data worse than the 
thermal model (represented by a cut-off power law, Section \ref{s:model}). 

The best fit of such a model to the \ginga/OSSE data shown by the solid curve 
in Fig.\ 3{\it a\/} corresponds to a low nonthermal compactness, $\ell=0.1$ 
(implying negligible \ee\ pair production), the compactness in blackbody 
photons 20 times the nonthermal compactness (assuring that the X\g\ spectrum 
is entirely due to the first-order Compton scattering), and the best-fit 
$\Gamma=1.85$. We see that the model is above the data at $\sim 150$--350 keV, 
which results in $\Delta\chi^2= +4$ with respect to the thermal model. 
Increasing the nonthermal compactness steepens the X\g\ spectrum (Svensson 
1987; Lightman \& Zdziarski 1987) and further worsens the fit. E.g., 
$\ell=20$, $\Gamma=1.8$ yield $\Delta\chi^2=+8$ with respect to the thermal 
model. Thus, we reject the ($\Gamma=1.8$)-injection model. 

On the other hand, Zdziarski et al.\ (1990) have proposed that the X-ray 
spectral index of $\alpha\simeq 0.9$--1 of Seyfert 1s is due to a nonthermal 
model with dominant pair production. The primary electrons are injected 
monoenergetically, which implies $\alpha=0.5$ in the absence of pair 
production (Blumenthal \& Gould 1970). Then, saturated nonthermal pair 
production yields the X-ray spectral index of $\alpha\simeq 1$ in the limit of 
$\ell\gg 10$ (Svensson 1987; Lightman \& Zdziarski 1987), which can 
explain the X-ray spectra of Seyfert 1s. However, this model also predicts a 
steepening of the spectrum at a few tens of keV due to downscattering of hard 
X-rays and soft \g-rays by thermalized, optically thick, pairs (Sunyaev \& 
Titarchuk 1980). This model was ruled out for IC 4329A by Z94, which 
conclusion we extend here to the average Seyfert-1 spectra. The dotted curve 
in Fig.\ 3{\it a\/} shows the best fit of this model ($\ell\simeq 30$). We see 
that the model lies below the $\sim 50$--150 keV spectrum, which results in an 
unacceptable $\Delta\chi^2 =+25$ with respect to the best-fit thermal model. 

However, we find that some nonthermal models intermediate between the two 
models rejected above fit our data satisfactorily (as also shown by Z94 for IC 
4329A). We consider here models with a power-law electron injection (between 
the Lorentz factors of 1.3 and $10^3$), pair production, and allowing for 
repeated Compton scattering. This yields $\Gamma= 5.0^{+4.4}_{-3.6}$ with the 
same $\chi^2$ as for the thermal model. The spectrum of the best-fit model 
($\ell\simeq 130$; the solid curve in Fig.\ 3{\it b}) is due to repeated 
Compton upscattering by both the nonthermal and the thermal parts of the 
electron/pair distribution (Zdziarski et al.\ 1990; Ghisellini, Haardt \& 
Fabian 1993). Note the similarity of the spectrum to that of the thermal model 
(Fig.\ 1). This nonthermal model yields no distinct pair-annihilation feature. 
The annihilation feature becomes stronger with decreasing $\Gamma$, 
illustrated by the dotted curve in Fig.\ 3{\it b\/} for $\Gamma=2.5$ 
($\ell\simeq 30$; $\Delta \chi^2=+1.6$). Note that the annihilation features 
in Fig. 3{\it a, b\/} are allowed the OSSE data and do not constrain the 
models by themselves. Decreasing $\Gamma$ also decreases the importance of 
repeated Compton upscattering. The spectrum for the lowest allowed $\Gamma$ 
($=1.4$) is dominated by the first-order nonthermal Compton scattering by both 
the injected nonthermal electrons and nonthermal pairs (which pairs steepen 
the X-ray spectrum from $\alpha= \Gamma/2=0.7$ to the observed value of 
$\alpha=0.9$). 

\begin{figure}
\label{fig:nth}
\begin{center}
\leavevmode
\epsfxsize=8.4cm \epsfbox{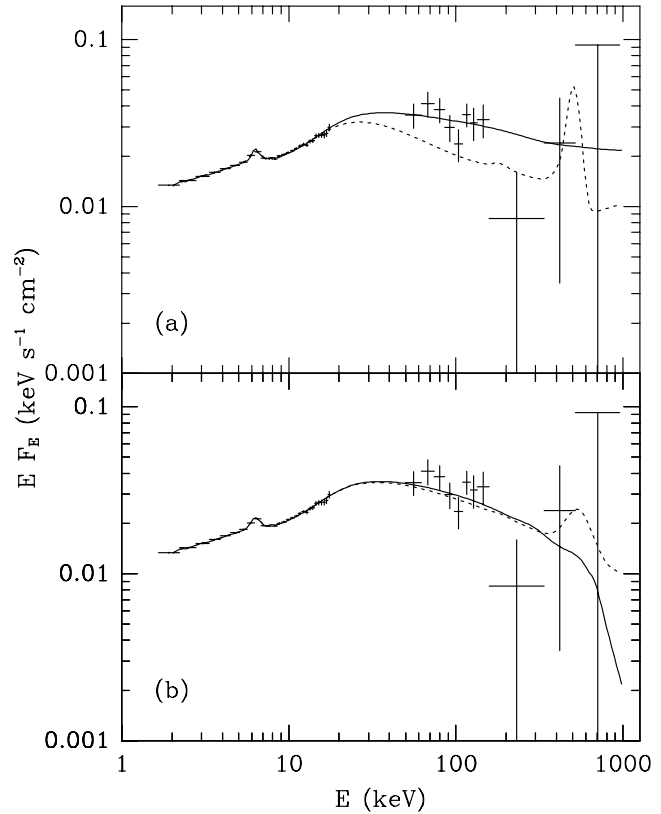}
\end{center}
\caption{
The average spectrum of RQ Seyfert 1s from \ginga\/ and OSSE modeled by 
nonthermal plasmas. {\it (a)} Two examples of rejected models. The solid curve 
corresponds to a model with power-law electrons injected with the index 
$\Gamma=1.85$ and with negligible both pair production and repeated Compton 
scattering. The solid curve corresponds to a model with monoenergetic 
injection of electrons and a saturated pair cascade (Zdziarski et al.\ 1990). 
{\it (b)} Nonthermal models with power-law injection of electrons, providing 
fits as good as the best-fit thermal model. The solid curve represents the 
best-fit model with $\Gamma=5.0$ and negligible pair production. The dashed 
curve represents a model with $\Gamma=2.5$ and moderate pair production. See 
Section 5.2. } 
 \end{figure}

\section{CONCLUSIONS}

We have obtained the average spectral parameters of RQ Seyfert 1s in X-rays 
and \g-rays based on samples of objects observed by both \ginga\/ and OSSE and 
by both \exosat\/ and OSSE. The estimates of the average spectrum from both 
samples are fully consistent with each other. Furthermore, they are consistent 
with the results of \heao\ A1 and A4. The average spectrum contains an 
underlying power law with $\alpha\simeq 0.9$. There is also a 
Compton-reflection spectral component corresponding to cold matter covering a 
solid angle of $\sim 1.5\pi$ as seen from the X\g\ source. The power law 
continues to soft \g-rays and it breaks or it is cut off with an $e$-folding 
energy of $\simgreat 250$ keV. The range of the $e$-folding energy obtained 
for our average spectra is fully consistent with that of IC 4329A (Z94; 
Madejski et al.\ 95). The average $e$-folding energy is {\it not\/} 40--50 
keV, which was reported earlier based on the OSSE data alone. 

Our average spectrum agrees well in X-rays with the average spectrum of all 
Seyfert 1s observed by \ginga\/ (NP94). The intrinsic dispersion in the 
spectral index and the contribution of reflection in Seyfert 1s is given by 
NP94. On the other hand, the limited statistics of our OSSE spectra allows us 
to provide only the range of the $e$-folding energy for the average spectrum 
with $\alpha=0.90\pm 0.05$. The value of the $e$-folding energy in 
individual AGNs appears to  correlate positively with the X-ray spectral index 
(Zdziarski et al.\ 1996). 

The average spectra of RQ Seyfert 1s can be modeled by Comptonization models 
with either thermal or nonthermal electrons. The Comptonizing plasma in 
thermal models is optically thin and relativistic. The plasma is likely to 
form a patchy corona above the surface of an accretion disk. Some models with 
relativistic nonthermal electrons predict the presence of an annihilation 
feature around 511 keV, which can be tested in the future by the {\it 
INTEGRAL\/} observatory (Winkler 1994). 

\section*{ACKNOWLEDGEMENTS}
This research has been supported in part by the Polish KBN grants 2P03D01008 
and 2P03D01410, NASA grants, and the NSF grant PHY94-07194. It has made use of 
data obtained through the High Energy Astrophysics Science Archive Research 
Center Online Service, provided by NASA/GSFC. We are grateful to L. Angelini, 
C. Done, and R. Svensson for valuable discussions, and R. Cameron for letting
us to use his data on NGC 7469 from OSSE. IMG acknowledges the financial 
support of the Universities Space Research Association.

\end{document}